\documentclass[%
 reprint,
superscriptaddress,
 amsmath,amssymb,
 aps,
floatfix,
]{revtex4-2}

\usepackage{mathrsfs,stmaryrd}
\usepackage[shortlabels]{enumitem}
\usepackage{dsfont}
\usepackage{multirow}
\usepackage{graphicx}
\usepackage{dcolumn}
\usepackage{bm}
\usepackage{hyperref}

\usepackage{pgfplots}
\pgfplotsset{compat=newest}
\usepackage{tikz}
\usetikzlibrary{automata,shapes.misc,positioning}
\usetikzlibrary{arrows,decorations.pathmorphing,bending}
\usetikzlibrary{calc}

\usepackage{xparse}
\DeclareDocumentCommand \esper { o m } {%
	\IfNoValueTF {#1} {%
		\ensuremath{\mathbb{E} \left[  #2 \right]} %
	}{%
		\ensuremath{\mathbb{E}_{#1} \left[  #2 \right]} %
	}%
}

\newcommand{\pars}[1]{\left(  #1 \right)}

\newcommand{\setof}[1]{\left\{ #1\right\}}

\newcommand{\rr}{\mathbb{R}}

\newcommand{\sqkets}[1]{\left[ #1 \right]}
\renewcommand{\d}{\mathrm{d}}

\setlist[description]{style=nextline}

\begin{document}

\preprint{APS/123-QED}

\title{Quantitative effects of the stress response \\to DNA damage in the cell size control of \textit{Escherichia coli}}

\author{Ignacio \surname{Madrid Canales}}
\email{madrid@sat.t.u-tokyo.ac.jp}
\affiliation{CMAP, \'Ecole polytechnique, CNRS, Institut polytechnique de Paris, Inria, Route de Saclay, 91120 Palaiseau, France}
\author{James \surname{Broughton}}
\email{j.broughton@ed.ac.uk}
\affiliation{University of Edinburgh, School of Biological Sciences, Institute of Cell Biology, UK}
\author{Sylvie \surname{M\'el\'eard}}
\email{sylvie.meleard@polytechnique.edu}
\affiliation{CMAP, \'Ecole polytechnique, CNRS, Institut polytechnique de Paris, Inria, Route de Saclay, 91120 Palaiseau, France}
\affiliation{Institut Universitaire de France}
\author{Meriem \surname{El Karoui}}
\email{meriem.elkaroui@ed.ac.uk}
\affiliation{University of Edinburgh, School of Biological Sciences, Institute of Cell Biology, UK}


\date{\today}

\begin{abstract}
In \textit{Escherichia coli} the response to DNA damage shows strong cell-to-cell-heterogenity. This results in a random delay in cell division and asymmetrical binary fission of single cells, which can compromise the size homeostasis of the population. To quantify the effect of the heterogeneous response to genotoxic stress (called SOS response in \textit{E. coli}) on the growth of the bacterial population, we propose a flexible time-continuous parametric model of individual-based population dynamics. We construct a stochastic model based on the ``adder" size-control mechanism, extended to incorporate the dynamics of the SOS response and its effect on cell division. The model is fitted to individual lineage data obtained in a 'mother machine' microfluidic device. We show that the heterogeneity of the SOS response can bias the observed division rate. In particular, we show that the adder division rate is decreased by SOS induction and that this perturbative effect is stronger in fast-growing conditions. 

\end{abstract}

\maketitle

\section{Introduction}
How cells control their size is a fundamental problem that has attracted much attention. The modelling, analysis and statistical calibration of this dynamic process, both from phenomenological and coarse-grained mechanistic approaches, has been studied by a large number of biologists, physicists and mathematicians \cite{Amir2014,Osella2014,Doumic2015,doumic2023individual}. Several variables and key checkpoint events have been proposed as candidates for drivers of cell division (see the reviews \cite{Meunier2021, Serbanescu2022} and the references therein). Nonetheless, the simple ``adder model", in which individual cells divide after adding a given amount of volume which is tightly controlled and uncorrelated to the initial cell size, has been shown to provide an excellent fit to the experimental distributions of \textit{Escherichia coli} cell sizes \cite{Taheri-Araghi2015}, in contrast to purely age-structured (``timer") or purely volume-structured (``sizer") models. The underlying molecular origins of the adder model have been explored recently, suggesting it is an emergent property of the coordination of DNA replication, RNA/protein allocation and protein accumulation \cite{Jun2015, Serbanescu2022}.

In \textit{E. coli}, the induction of DNA damage triggers a complex molecular response called the \textbf{SOS response} \cite{Witkin1967,Little1982}, which is essential for repair. Recent developments in high-throughput single-cell imaging techniques have allowed the observation of the effect of the SOS response on an individual scale \cite{Friedman2005,Jones2021}. Among these, the microfluidic device called the mother machine (MM) \cite{Wang2010, Ollion2019}, is designed to track multiple single bacteria over several generations, thus permitting long-term continuous imaging \cite{jaramillo22}. In particular, the intensity of the stress response can be monitored over time using fluorescent transcriptional reporter markers. These experiments have revealed significant heterogeneity in the SOS response among individual cells \cite{jaramillo22,Jones2021,diaz-diaz_heterogeneity_2024}, ranging from very weak to very strong, with substantial impacts on their morphology and growth \cite{Ojkic2020}. Indeed, the SOS response induces the expression of proteins that cause cell division to stop without arresting cell growth. Consequently, the SOS heterogeneity is translated into the emergence of a subpopulation of abnormally long ``filamentous" bacteria \cite{wehrens2018,jaramillo22}. 

Although previous experiments have suggested that the adder model is robust under diverse kinds of growth inhibitions \cite{Si2019}, the multifactorial effects of the SOS response may lead to perturbations of the adder model, especially for filamentous bacteria. Interestingly, once the stress is removed, filamentous bacteria can resume proliferation through a series of asymmetrical divisions \cite{Raghunathan2020,wehrens2018,Cayron2023}. Moreover, their divisions are known to fulfil, on average, the adder hypothesis \cite{wehrens2018,Raghunathan2020}. However, it is unclear how robust the adder model is under such a heterogeneous response and how fast it can restore size homeostasis upon exposure to an antibiotic that causes DNA damage. 

Here, we introduce a parametric time-continuous model of cell proliferation that takes into account the SOS response. We then fit this model to single-cell lineage data acquired in the mother machine where cells were exposed to the antibiotic ciprofloxacin at sublethal levels \cite{JamesBroughton2024DATA,Broughton2024}. We show that our model explains the observed dynamics and allow us to give quantitative insights about the effect of the SOS response over the division mechanism under different nutrient conditions. We show first that an Ornstein-Uhlenbeck model for the expression level of the SOS response recovers the population distributions when the antibiotic is added and subsequently removed from the medium. Second, we show that the adder division rate is decreased by SOS induction and that this perturbative effect is stronger in fast-growing conditions. Finally, we show that the observed asymmetrical divisions of filamentous bacteria can be well-fitted by a Beta mixture model, which shows a larger variance in fast-growing media.

	\section{A time-continuous single-cell adder model under stress}
	\label{sec:modelAdder}
	We consider a stochastic formulation which accounts for the individual (single-cell) variability within the population. To that extent, each individual cell $i$ is characterised by a three-dimensional vector $\boldsymbol{\xi}_i(t) = (a_i(t), y_i(t), x_i(t))$ consisting of:
	\begin{itemize}
		\item $a_i(t)$, its \textbf{added size} from birth to current time $t$,
		\item $y_i(t)$, its \textbf{current size} at time $t$, and
		\item $x_i(t)$, the \textbf{SOS level} (fluorescence) at time $t$.
	\end{itemize} 
	We assume that each cell behaves independently. The population then evolves in continuous time through three fundamental dynamics: stress response, growth, and division. While growth is assumed to be deterministic, the division and stress response mechanisms will account for the observed stochasticity. We suppose that the stress response is independent from the growth dynamics, while, on the other hand, the division mechanism is affected by the level of stress response. 

\begin{figure}[t]
	\includegraphics[width=0.45\textwidth]{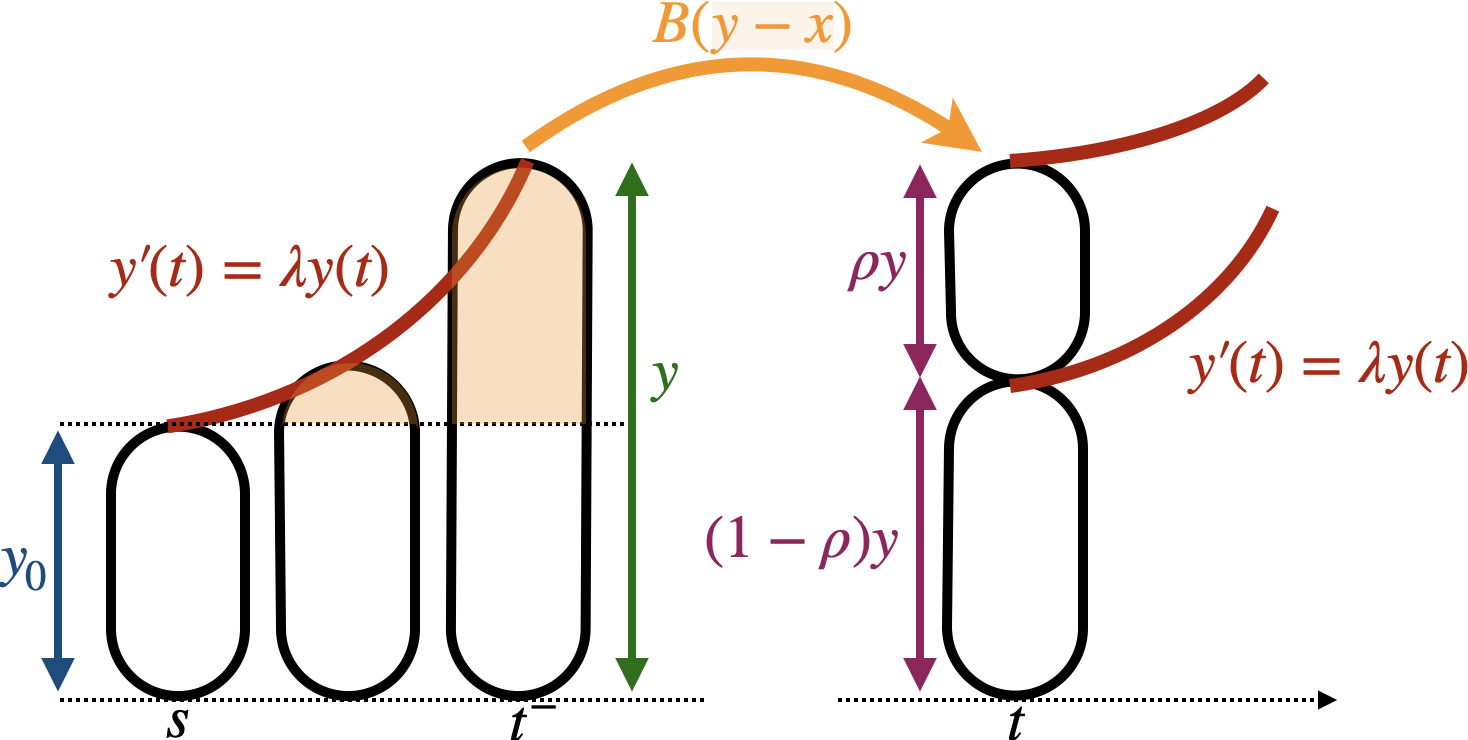}
	\caption{\textbf{The adder model of cell division}}
		\label{fig:AdderModel}
	\end{figure}

 In particular, we consider the case where the antibiotic inducing the SOS response is present only between times $\tau_0 < t \leq \tau_1$. The continuous-time model will allow us to study the transient phases of the introduction of the antibiotic at $t = \tau_0$ and of its removal at $t = \tau_1$. The model is detailed below and summarised by Fig. \ref{fig:AdderModel}.

\subsection{SOS response}

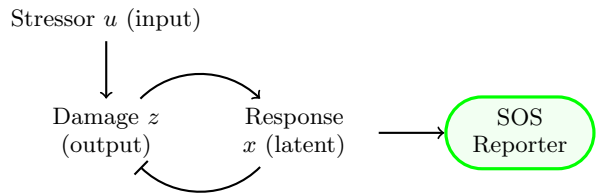
\begin{figure}[h]
		\centering
	\begin{tikzpicture}[font=\small,thick]
	\node (A) at (-1,1.5) [text width=3cm,align=center]{Stressor $u$ (input)};
	\node (B) at (-1,0) [text width=2cm,align=center]{Damage $z$ (output)};
	\node (C) at (1.5,0) [text width=2cm,align=center]{Response $x$ (latent)};
	\node[draw,rounded rectangle, draw=green, fill=green!5, very thick] (gfp) at (4.5,0) {\begin{tabular}{c} SOS \\ Reporter \end{tabular}}; 
	\draw[->] (A) edge (B)
	(B) edge [bend left=45] (C);
	\draw[-{Bar}] (C) edge [bend left=45]  (B);
	\draw[->] (C) edge (gfp);
	\end{tikzpicture}
	\caption{Scheme summarising the negative autoregulation models of the SOS response of \cite{Szekely2013}. Arrows marked $\to$ represent positive regulation (v.g. synthesis or disinhibition), while $\dashv$ represents negative regulation (repression or inhibition). The fluorescent SOS appears highlighted.}
	\label{fig:NARscheme}
\end{figure}

Previous simulation studies that have looked at the expression levels of several proteins participating in the SOS response, have shown that the SOS regulatory network can be accurately modelled by low-dimensional chemical reaction models \cite{Friedman2005,Shimoni2009,Szekely2013}. These models are typically characterised by negatively autoregulated motifs, as shown in Fig. \ref{fig:NARscheme}, where a stressor $u$ produces some damage $z$, which triggers a response $x$ that, in turn, repairs the damage $z$. 

 To account for this dynamic feedback, the authors in \cite{Szekely2013} propose a simple deterministic model they name integral feedback model. It supposes that the amount of damage $z(t)$ is the result of the difference between the value of the stressor signal $u(t)$ (damage induction) and the stress response $x(t)$ (damage repair). At the same time, the stress response $x(t)$ senses the damage, and its intensity increases linearly with the level of damage $z(t)$ with a proportionality factor equal to $\theta > 0$. The parameter $\theta$ thus represents the rate of reactivity of the stress response with respect to the perceived damage.

Here, we generalise the deterministic integral feedback model to account for the fact that the SOS response is not coordinated and varies significantly between individuals and in time. We propose to model the dynamics of the SOS signal as a real-valued process $(X_t)_{t\geq 0}$ solution to the Orstein-Uhlenbeck Stochastic Differential Equation
    \begin{equation}
    \d  X_t  = \  \theta_{c(t)} (\mu_{c(t)} - X_t) \d t + \zeta_{c(t)} \d B_t
    \label{eq:OU}
    \end{equation}
    where $B_t$ is a standard Brownian motion, $c(t) = \mathds{1}_{t \in [\tau_0,\tau_1[}$ equals 1 whenever the antibiotic is present in the medium and 0 otherwise, and $\theta_i, \mu_i, \zeta_i$, $i \in \{0,1\}$ can be interpreted as follows:
    \begin{itemize}
        \item $\mu_i$ is the basal SOS expression level under stress $i \in \{0, 1\}$.
        \item $\theta_i > 0$ measures the strength at which the SOS expression reverts to its basal level after periods of under or over-expression. It is related to the molecular rates of induction and repression of the SOS response.
        \item $\zeta^2_i>0$ is the variance with which the log-fluorescence level fluctuates around the average value, accounting for various potential sources of stochasticity in the signal. 
    \end{itemize}

	\subsection{Growth} 
	Each cell $i$ of size $y_i$ grows exponentially \cite{Godin2010} at elongation rate $\lambda > 0$ which we assume to be the same for the whole population:
	\[ \frac{dy_i(t)}{dt} = \lambda y_i(t). \]
	Thereby, the size $y(t)$ and added size $a(t)$ at time $t \geq 0$ of a bacterium which had size $y(s)$ and added size $a(s)$ at time $s<t$ are given by the following deterministic equations:
	
	\begin{align}
	y(t) &= y(s) \exp(\lambda (t-s))
	\label{eq:fluxY} \\
	a(t) &= a(s) + y(s) \exp(\lambda (t-s)) - y(s)
	\label{eq:fluxa}
	\end{align}

	\subsection{Division} 
 \label{sec:division}

 We suppose that divisions occur at random times, driven by an instantaneous division rate $\beta \geq 0$ that depends on the current cell trait $\boldsymbol{\xi}_t = (a_t, y_t, X_t)$ such that
    \begin{align}
    \mathbb{P}(\textrm{Division time} \in [t+\d t[ \ |& \textrm{Division time} \geq t, \boldsymbol{\xi}_t )  \nonumber \\
    &= \beta(\boldsymbol{\xi}_t) \d t + o (\d t).
    \label{eq:betaAYS}
    \end{align}
    Within the framework of the unperturbed adder model \cite{Taheri-Araghi2015,gabriel:hal-01742140} (forgetting the SOS level $X_t$), a bacterium of birth size $y_0$ will divide at a random size $y_0 + A_{div}$ where the added size $A_{div}$ is independent from $y_0$, distributed according to
	\begin{equation}
	S(a) = \mathbb{P}(A_{div} \geq a) = \exp \left( - \int_0^a B(\alpha) d\alpha \right).
	\label{eq:S(a)}
	\end{equation}
 In this case, $\beta$ is written as
    \begin{equation} 
    \beta(a,y,x) = \lambda y B(a) ,
    \label{eq:beta_adder}
    \end{equation}
    where $B$ is the adder division rate and $\lambda$ is the elongation rate. Under the adder hypothesis, the function $B$ depends only on $a$.  To incorporate the effect of SOS response, we propose rates $B$ that depend also on the value of the SOS level $x$. Notice that in general this will result in the lost of the adder property. Indeed, since the SOS response evolves through time, its level at division depends on the total duration of the interdivision time, which depends on the birth size for the adder model. As middle ground, we consider the case in which the SOS level affects the fluctuations of the adder around its ideal value, rather than the added amount of size itself. To that extent, we consider the case where conditionally to the SOS level $x$, the division rates $B(a,x)$ are those of a Generalised Gamma distribution \cite{Stacy1962, Prentice1974}, detailed in Appendix \ref{app:gengamma}, with scale parameter independent of $x$, and shape parameters depending linearly on the value of $x$ through two values $s_1$ and $q_1$ to be estimated (see the Appendix for their definition). 
    
\smallskip

Finally, when a division occurs, we let $\rho \in [0,1]$ the ratio between the observed size of the followed mother after and before division and we write $k(y,\rho)$ the probability density that a mother of size $y$ produces a daughter of size $\rho y$. The mother machine allows to identify the progenitor cell for each followed individual. This allows to compute the ratio $\rho$ between the birth size of each cell and the division size of its mother. The experimental results and PDE simulations carried out by Wehrens et al. \cite{wehrens2018} show that the division septa appear around very precise positions determined by the nodes of the stationary solutions of a reaction-diffusion model of proteins (the Min system). Moreover, they show that the number of these positions increases with the mother size $y$. In concrete, they show that both the empirical and simulated distributions of $\rho$, reach their maximum at constant positions $(w_n^N)_{n \in \llbracket 1, N \rrbracket} \in ]0,1[$ determined by the total number $N$ of nodes and given explicitly by 
    \begin{equation}
    w_n^N = \frac{2n-1}{2N}.
    \label{eq:wnM}
    \end{equation}
    
    Based on these previous findings and our own observations, we propose a parametric model for $k(y,\rho)$. First, we call $N(y)$ the number of possible septa, which is function of the mother size $y$. In particular, we suppose that there is a critical size parameter $y^*$ that determines the number of possible septa by the rule
    \begin{equation}
    N(y) = \sqkets{\frac{y}{2y^*} } + 1 ,
    \end{equation}
    where $[u]$ is the integer part of $u$. This means that if the mother size is below $2y^*$ there is only one possible septum ($N = 1$), if the mother size is between $2y^*$ and $4y^*$, then there are two possible septa ($N =2$), etc. As such, $y^*$ can be thought as related to the characteristic wavelengths of the standing waves produced by the Min system. 
    
    \smallskip
    
    Additionally, we suppose that, conditionally to the mother size $y$, the division can occur at each possible septum with equal probability, this is, each possible site can be chosen uniformly with probability $1/N(y)$. Therefore, we suppose that $k$ is of the form
    \begin{equation}
    k(y,\rho) = \frac{1}{N(y)} \sum_{n=1}^{N(y)} F_{n}^{N(y)} \pars{\rho} ,
    \label{eq:kSOS}
    \end{equation}
    where, for all fixed $N$ and $n$, $F^N_n(\rho)$ is the probability density of producing a daughter of size $\rho$ times the size of the mother, when the division happens at the $n$-th septum among the $N$ possible ones. 
    
    \smallskip
    
    Finally, we suppose that cells have no particular orientation. For example, if $N = 3$, the first and the third septa cannot be distinguished. In general, this imposes that our densities $F$ have to verify, for all $n \leq N/2$ and all $\rho \in [0,1]$ a symmetry condition written
    $$
    F^N_n(\rho) = F^N_{N+1-n}(1 - \rho).
    $$
    In particular, we suppose that $F^{N}_n$ is the probability density function of the Beta distribution of parameters $(\alpha_n^N, \beta_n^N)$. This is
    \begin{equation}
    F^N_n(\rho) = \frac{\Gamma (\alpha_n^N + \beta_n^N) }{\Gamma (\alpha_n^N) \Gamma( \beta_n^N)} \rho^{\alpha_n^N  - 1 } (1 - \rho)^{\beta_n^N -1 }.
    \label{eq:kSOS_Beta}
    \end{equation}
    The findings of Wehrens et al. suggest to take, for all $N>1, n \in \llbracket 1, N\rrbracket$, the value of $w_n^N$ defined in \eqref{eq:wnM} as the mode of the distribution $F_n^N$. This is $w_n^N = \textrm{argmax}_{0 \leq \rho \leq 1} F_n^N(\rho)$ (i.e. as the \textit{peak} of the observed distribution). Since for $\alpha_n^N > 1$ and $\beta_n^N > 1$ the mode is given by $\frac{\alpha_n^N-1}{\alpha_n^N + \beta_n^N - 2}$ (otherwise equal to 0 and 1, and therefore not of our interest), we want
    \[
    w_n^N = \frac{\alpha_n^N-1}{\alpha_n^N + \beta_n^N - 2}.
    \]
    Let the denominator be called $v_n^N := \alpha_n^N + \beta_n^N -2 > 0$. Then we can write
     \begin{align*}
        \alpha^N_n &= 1 + v_n^N w^N_n, \\
        \beta^N_n &= 1 + v_n^N \pars{ 1 - w^N_n }.
    \end{align*}
    As the next computation shows, the variance of $\rho$ is a decreasing function of $v_n^N$:
    \begin{align*}
    \textrm{Var}(\boldsymbol{\rho}) &= \frac{\alpha_n^N \beta_n^N}{(\alpha_n^N + \beta_n^N+1)(\alpha_n^N + \beta_n^N)^2} \\ 
    &= \frac{1 + v_n^N + (v_n^N)^2(w_n^N - (w_n^N)^2)}{(v_n^N+3)(v_n^N+2)^2}.
    \end{align*}  
    
    As such, $v_n^N$ is an inverse measure of the dispersion of the distribution of $\rho$. We make the biological assumption that the concentration of division proteins around the chosen septum is independent of the length of the cell and the total number of possible septa. This translates as setting for all $N \geq 1$, $n \in \llbracket 1, N \rrbracket$, $v_n^N = v > 0$ constant, depending only on the culture medium. 
    
    Therefore, using the parametrisation, the kernel $k$ depends only on two parameters (for each medium): the critical size $y^*$ which defines the number of possible septa, and the constant $v$ which determines the dispersion of the Beta distributions around them. 

\section{Results}
\label{sec:7_SOSadderSimulation}

From now on, we call $\eta_X$ the parameters associated with the Ornstein-Uhlenbeck model of SOS expression, $\eta_{\beta}$ the parameters associated with the Generalised Gamma Adder model of the division rate, and $\eta_{k}$ the parameters associated with the cell division kernel. Using the procedure presented in Appendix \ref{app:MLE}, we estimate the most likely values of $\eta = (\eta_X, \eta_\beta, \eta_k)$, which allow us to carry the simulations presented in the following section. The data considered consist on observations under three different growth media: fast-growing (M9 minimal salts supplemented with glucose and amino acids), intermediate growing rate (M9+glucose) and slow-growing (M9+glycerol), where 3 $ng/ml$ of the antibiotic ciprofloxacin is added to the medium between hours $t = 2$ and $t = 14$ of the experiment, which last for a total of $T = 25$ hours. See \cite{Broughton2024} for the methodological details and \cite{JamesBroughton2024DATA} for the raw datasets.

\subsection{The Ornstein-Uhlenbeck model recovers the population distributions and the induction and recovery phases of the SOS response}
Table \ref{tab:SOS_MLEvalues} gives the estimated parameters of the Ornstein-Uhlenbeck model for the SOS response dynamics. The estimated parameters show that the presence of a sublethal concentration of ciprofloxacin produces a shift in the mean value $\mu$ of the SOS signal of 10-100 times the basal fluorescence (the values in the table are in log-scale), without significantly changing the regulation parameter $\theta$. This suggests that ciprofloxacin leads to an increased production of the fluorescent marker measured, without altering the average response rates of the SOS regulatory circuit itself. That is without affecting the turning off of the SOS response once the damage is repaired, for example. At the same time however, the noise of the SOS intensity, conveyed by $\zeta$, is systematically superior under exposure to ciprofloxacin, which could indicate the presence of additional perturbations. Nonetheless, the small variations of these parameters point towards a certain robustness of the SOS response. With respect to the the variations in different media, we observe that the SOS dynamical response is noisier under fast-growing conditions and exhibits larger values of $\theta$. In particular, this means that the steady state of SOS intensity is reached faster in fast-growing media which is expected as the response rate of bacterial systems depends largely on their growth rate. On the other hand, the mean SOS expression is slightly increased in intermediate and slow-growth media compared to fast growht, as expected from the bacterial growth laws.

\begin{table}[h]
    \centering
    \begin{tabular}{|c|c|c|c|c|c|c|}
    \hline
         \multirow{3}{*}{\textbf{Medium}} & \multicolumn{6}{c|}{\textbf{SOS diffusion parameters}} \\ \cline{2-7}
         & \multicolumn{2}{c|}{$\theta_c$} & \multicolumn{2}{c|}{$\mu_c$} & \multicolumn{2}{c|}{$\zeta_c^2$} \\
         \cline{2-7}
         &  Cip$-$ & Cip+ &   Cip$-$ & Cip+  &   Cip$-$  & Cip+ \\
         \cline{1-7}
         gly   & 0.217 & 0.233 & 4.57 & 6.01 & 0.0702 & 0.103 \\
         glu   & 0.319 & 0.223 & 4.60 & 6.12 & 0.1074 & 0.117 \\
         gluaa & 0.423 & 0.351 & 4.53 & 5.46 & 0.1439 & 0.161 \\
         \hline
    \end{tabular}
    \caption{Maximum Likelihood Estimators of the parameters driving the Ornstein-Uhlenbeck Equation \eqref{eq:OU} for the three different media, and under the presence or not of ciprofloxacin (log-scale of fluorescence measurements (arbitrary units)).}
    \label{tab:SOS_MLEvalues}
\end{table}

\begin{figure*}[t]
    \centering
    \includegraphics[width=0.7\textwidth]{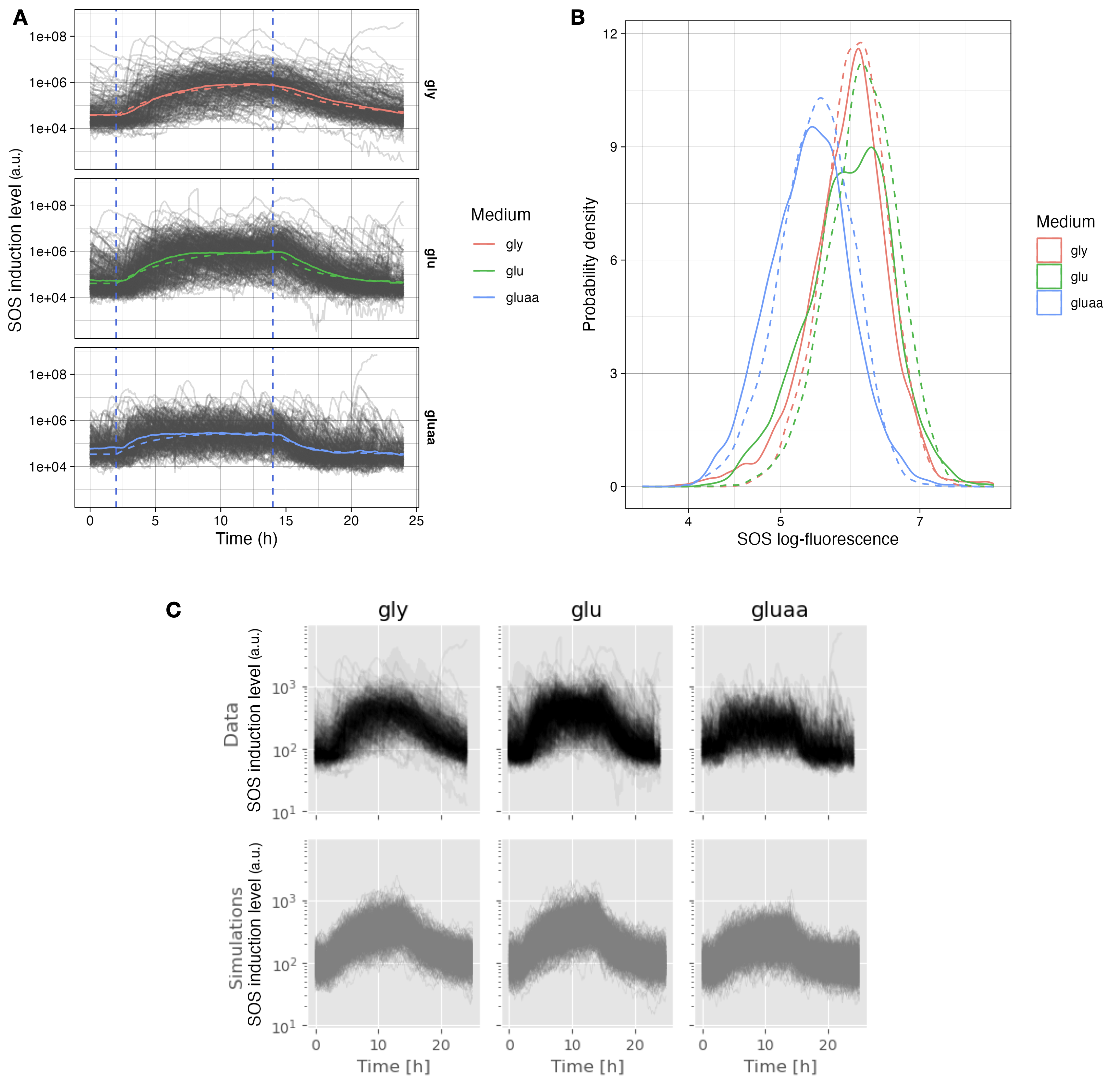}
    \caption{\textbf{Fitting results of the SOS dynamics using the Ornstein-Uhlenbeck model: dynamical and stationary.} \textbf{A.} In solid lines, empirical mean of the SOS intensity calculated from the grey trajectories in the background, which represent individual cells. Only lineages with observed divisions were retained. In dashed lines, the mean SOS intensity predicted by Eq. \eqref{eq:OU} using the parameters summarised in Table \ref{tab:SOS_MLEvalues}. The time window of the ciprofloxacin treatment is delimited by vertical dashed lines. \textbf{B.} In solid lines, empirical steady-state distribution of the SOS intensity under ciprofloxacin, obtained from the data observed between $t = 10$ and $t = 14$ (last 4 hours of treatment, 8 hours after initial dose). In dashed lines, stationary distribution expected from the Eq. \eqref{eq:OU} using the parameters summarised in Table \ref{tab:SOS_MLEvalues}. \textbf{C.} Comparison of the empirical fluorescence observations (first row) and the simulated trajectories of \eqref{eq:OU} using the MLE.}
    \label{fig:SOSfits}
\end{figure*}

The SOS predictions of Eq. \eqref{eq:OU} using the parameters $\eta_X$ given by Table \ref{tab:SOS_MLEvalues} are shown in Fig. \ref{fig:SOSfits}. Panel A of the figure shows the predicted mean SOS intensity over time (dashed line) and compares it to the empirical mean (solid line). We see an excellent qualitative agreement between the two curves. The fast regulation dynamics conveyed by the Ornstein-Uhlenbeck, which models a rather instantaneous regulation, seem to capture well the dynamical transition at times $t = 2$ and $t = 14$. In particular, the model recovers very accurately the shape of the inflexion observed at this regime change. We see also that, as already discussed above and as is shown by the values of $\theta$ in Table \ref{tab:SOS_MLEvalues}, the steady-state is reached faster in fast-growing media.

Panel B of Fig. \ref{fig:SOSfits} shows the empirical (solid line) and predicted (dashed line) steady-state distributions of SOS intensity under the effect of ciprofloxacin. As the Ornstein-Uhlenbeck process \eqref{eq:OU} is stationary, the predicted distribution is given explicitly by its stationary distribution: a Gaussian of mean $\mu$ and variance $\zeta^2/(2\theta)$. We also see an excellent qualitative agreement in the steady-state distribution, particularly around the mean value, except for the glucose medium, whose distribution is wider than the predicted Gaussian. In general, the observed distributions are more skewed to the left. This can be explained by the fact that the Ornstein-Uhlnebeck process imposes a symmetrical noise around the mean value, while the observations show that cells, even under ciprofloxacin, tend to concentrate below the expected value. At the same time, but more rarely, some lineages can induce very strongly the SOS response, which also widens the distribution towards the right. 
In panel C we compare the empirical  distribution (top row) with the simulations (bottom row). Here again, the OU process captures very the qualitative behaviour, but we notice some outlier cell lineages that are not recovered in the simulations, suggesting that there is additional noise that is not accounted for by the OU fluctuations. Taken together, these results indicate that the Ornstein-Uhlenbeck process captures well the dynamics of the SOS response under sub-lethal levels of ciprofloxacin.

\subsection{The adder division rate is decreased by SOS induction and its perturbative effect is stronger in fast-growing conditions}
To understand the coupling of the SOS level and the division rate, we calculated the Maximum Likelihood Estimators of the parameters of the Generalised Gamma model introduced in Appendix \ref{app:gengamma}. As seen in Table \ref{tab:Division_MLEvalues} (right part), in all three media, the multiplicative noise effect conveyed by $s_1$ is much less significant than the shape change effect conveyed by $q_1$. Moreover, $q_1$ increases with the richness of the medium. This shows that the perturbative effect of the SOS response on the adder control is stronger in fast-growing conditions. 

\begin{figure*}
    \centering
    \includegraphics[width=0.9\textwidth]{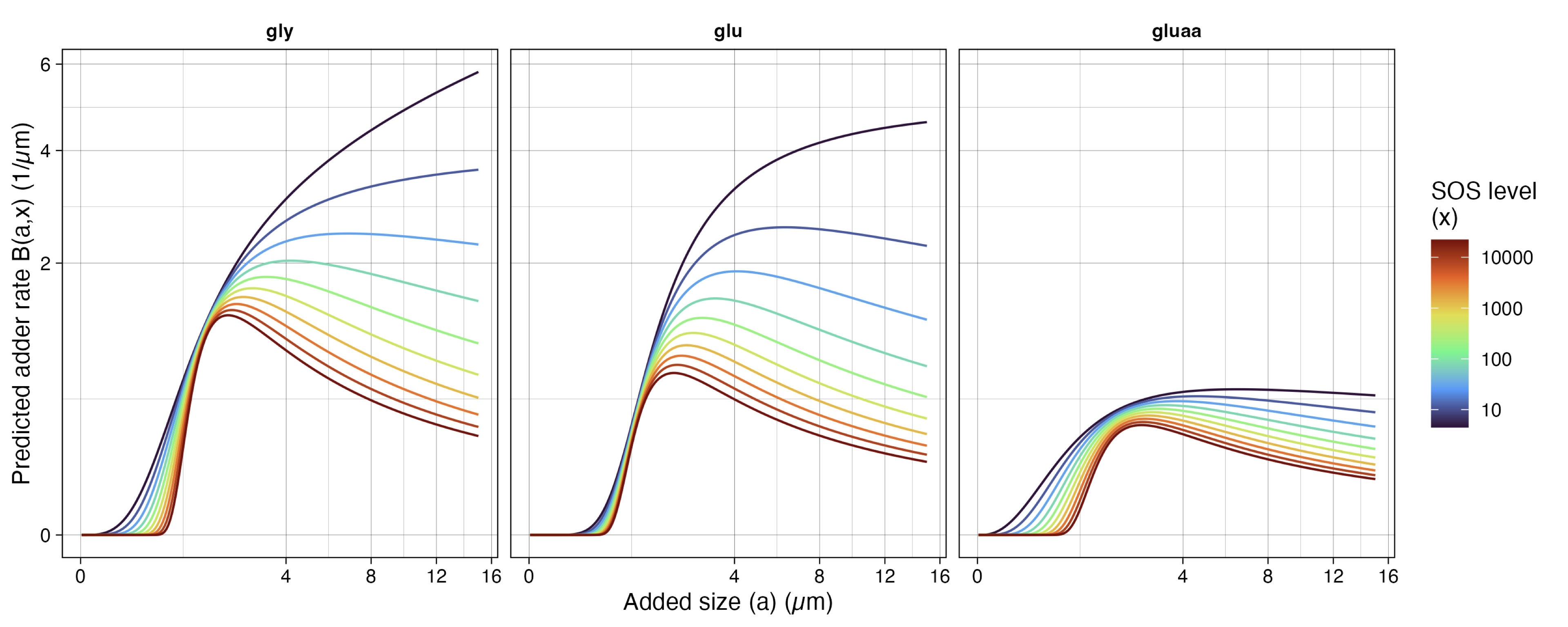}
    \caption{\textbf{Predicted adder division rate $B(a,X)$ by MLE under the three different media}}. The adder rate as a function of added size $a$ (abscissa) and SOS level $X$ (colors) predicted from the fitted parameters of Table \ref{tab:Division_MLEvalues} for SOS levels varying from very low (10, blue) to very high (10000, red) in the three growth conditions. 
    \label{fig:predictedB}
\end{figure*}

To observe more clearly the effects of $s_1$ and $q_1$ on the division dynamics, we calculate the division rates $B(\cdot,x)$ predicted by the MLE for different values of $x$ in all three media, as shown in Fig. \ref{fig:predictedB}. First, we find the expected result that the division rate is lower in richer media (that is, divisions occur fast in rich media) regardless of the SOS level. Second, we observe a strong impact of the SOS level. In all three media, increasing the SOS level leads to lower division rates, in keeping with the fact that the SOS response induces a delay in division. The strength of this inhibition seems to be stronger in poorer media, where the division rates in the absence of stress are higher. In other words, it seems that the adder size control is more sensitive to stress in poorer growth conditions. Not only is the value of the division rate changed, but so is the shape of its dependence on the added size. At low SOS levels, in all three media, the most likely division rate is an increasing function of the added size. Indeed, the intercept parameters $s_0$ and $q_0$ (see Table \ref{tab:Division_MLEvalues}) are all three in the ``increasing" region of the parameter space represented in Fig. \ref{fig:gengammas_cartoon}. However, as $x$ increases (i.e. the SOS level gets higher), the division rate changes its shape and tends in all three media towards an arc-shaped ``lognormal-like" distribution, similar to the path $A$ followed at the right panel of Fig. \ref{fig:gengammas_cartoon}. In these conditions, cells that have increased their length by a large amount are less likely to divide than shorter cells. This apparent contradiction emerges from the fact that when we measure the division rate as a function solely of the added size, we implicitly marginalise over all the unknown individual variables that might have an effect. This is similar to the problem of \textit{random effects} in the Mixed Effect Models literature \cite{Lavielle2014}. Indeed, if starting from $B(a,x)$ we wanted to obtain a \textit{real adder} division rate $\bar B(a)$, function of $a$ only, we can compute
\[
\bar B(a) :=  \frac{\d}{\d a} \pars{ - \log \bar S(a) }, 
\]
where $\bar S$ is the \textit{population survival function} given by
\[
\bar S(a) = \mathbb{P} \pars{A_{div} \geq a} = \esper{ \exp \pars{ - \int_0^{a} B(a,X_{\tau(a)})} \d a},
\]
where the expectation is taken over the Ornstein-Uhlenbeck process $X$ and $\tau(a) = \lambda^{-1}\log (1+y_0^{-1} a)$ is the time needed to reach an added size $a$ given the birth size $y_0$. Hence, under suitable integrability assumptions for $\bar S$ we can differentiate under the expectation sign and then
\begin{align*}
\bar B(a) &= \frac{\esper{B(a, X_{\tau(a)}) e^{- \int_0^{a} B(s,X_{\tau(s)}) \d s }  } }{\esper{- \int_0^{a} B(s,X_{\tau(s)}) \d s } } \\
&= \esper{\left. B(a, X_{\tau(a)}) \right| A_{div} \geq a} \neq \esper{ B(a, X_{\tau(a)}) }.
\end{align*}
This relation evidences a bias on $\bar B(a)$. Only the individuals that have yet not divided at added size $a$ contribute to the value of $\bar B(a)$. Therefore, if only cells with very high SOS intensity $x$ survive until longer added sizes, and the conditional division rate $B(a,x)$ is lower for high $x$, then the marginal rate $\bar B(a)$ will be lower for larger $a$, producing an ``effective catastrophe" region in $\bar B$, as for example observed by \cite{Osella2014} in a size-structured model, even if $B(a,x)$ were not decreasing themselves. In our case, if the stress is low enough, division rates are monotonically increasing (as one naively might expect for a homeostatic system: the more size that has been added, the more likely should the cell divide). And, if the stress is high enough they tend towards a characteristic arc-shaped division rate. In other words, our findings suggest that it is the high-SOS cells that lead to an apparent depression of the division rate when measuring it with population-level statistics.

\subsection{The Beta mixture model recovers the observed asymmetrical divisions of filamentous bacteria, with larger fluctuations in fast-growing media}
To analyse the division dynamics, we computed the MLE of $y^*$ and $v$ as shown in Table \ref{tab:Division_MLEvalues}. The value of MLE of $y^*$ is close to the mean division size observed in the empirical control dataset (3.25 $\mu$m in \textit{gly}, 3.53 $\mu$m in \textit{glu}, and 4.89 $\mu$m in \textit{gluaa} \cite{Broughton2024}). This confirms the interpretation of $y^*$ as a characteristic length of un-perturbed bacteria. The parameter $v$, which measures the inverse of the dispersion of the position of the septum is higher in the poorest medium. That is to say, the septum position is less precise in rich nutrient conditions. This confirms the trend observed with no antibiotic (see Appendix \ref{app:fitF_ctrl}), where the septum position seemed also to be less precise in fast-growing media. 

\begin{table}
    \centering
    \begin{tabular}{|c|c|c|c|c|c|c|}
    \hline
         \multirow{3}{*}{\textbf{Medium}} & \multicolumn{6}{c|}{\textbf{Division parameters}} \\ \cline{2-7}
         & \multicolumn{4}{c|}{Adder rate $\beta$} & \multicolumn{2}{c|}{Septum kernel $k$}\\
         \cline{2-7}
         &  $s_0 \dag$ & $s_1$ & $q_0 \dag$ & $q_1$ &   $y^*$  ($\mu$m) & $v$  \\
         \cline{1-7}
         gly   & 0.4207 & -0.009473 & 0.7832& -0.1552 & 3.6896 & 122.758 \\
         glu   & 0.3019 & 0.01579 & 0.5559 & -0.1552 & 3.6896 & 127.586  \\
         gluaa & 0.6626 & -0.007368 & 0.6963 & -0.2026 & 5.7586 & 93.7931 \\
         \hline
    \end{tabular}
    \caption{Maximum Likelihood Estimators of the parameters driving division for the three media.. The adder division rate $B$ depends on the values of $s_0$, $s_1$, $q_0$ and $q_1$ as defined in Appendix \ref{app:gengamma}. The mother-to-daughter ratio kernel $k$ depends on the characteristic length $y^*$ and dispersion parameter $v$ introduced in Section \ref{sec:division}. Parameters marked by $\dag$ are inferred by fitting a Generalised Gamma distribution directly to the added size distributions of the first SOS decile of control cells.}
    \label{tab:Division_MLEvalues}
\end{table}

\begin{figure*}[t]
    \centering
    \includegraphics[width=0.8\textwidth]{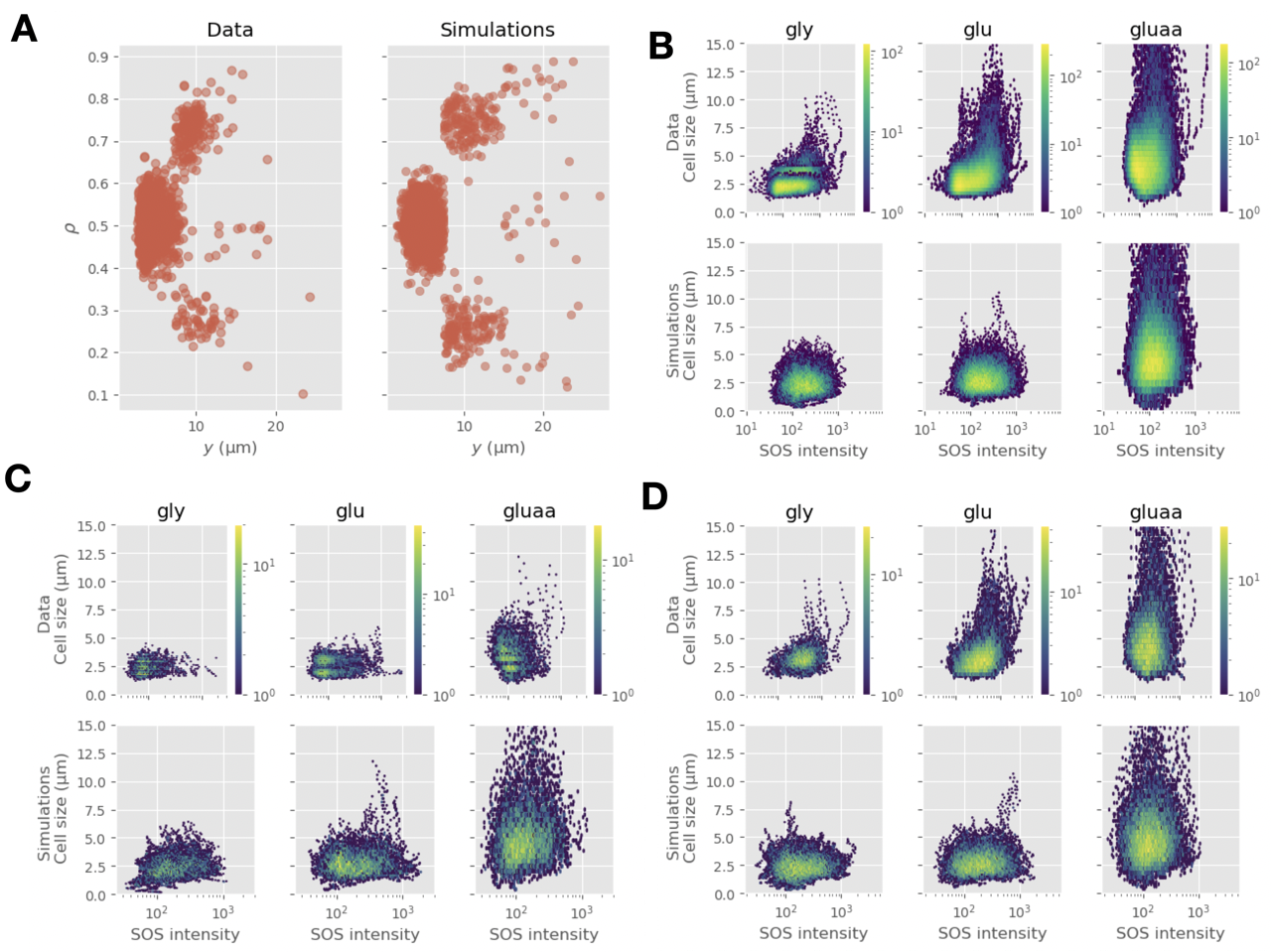}
    \caption{\textbf{A.} Data and simulations of $\rho \sim k(\cdot | y)$ (septum position given the mother size) for $I = 100$ independent lineages and the MLE as parameters (glucose+aa medium, where a larger number of possible septa can be observed). \textbf{B-D.} Data and simulations of the joint distributions of the SOS level at division and the division size. \textbf{B} is the time average over the whole experiment, \textbf{C} is during the first 2 hours, and \textbf{D} during the last 4 hours of ciprofloxacin exposure ($10 \leq t < 14$).}
    \label{fig:datasimu}
\end{figure*}

Fig. \ref{fig:datasimu}A shows that the division statistics are well recovered by our Beta mixture model. However,  the transition boundaries determined by $y^*$ are less marked in the empirical observations, suggesting the presence of individual heterogeneity that our model does not fully capture. Finally, Fig. \ref{fig:datasimu}B-D shows the joint distributions of size and SOS response averaged over various time windows. The bulk of the distribution seems well recovered, particularly during the presumed stationarity reached after 12 hours of antibiotic treatment (Panel D). However, some rare events associated to \textit{excessive} filamentation seem not be captured by the model. This also might indicate the presence of individual heterogeneity in the parameters of the division rate, which cannot be explained only by the SOS measurements. 

\section{Conclusions}
\label{sec:8_Conclusion}

We have proposed a parametric model of the perturbative effects of ciprofloxacin-induced SOS response over the adder model of size control in \textit{E. coli} in several growth conditions.  Our findings coincide with the previous observations that the adder model is robust to growth modulation \cite{Si2019}. In contrast, we have found that the SOS response, which is known to have multifactorial physiological effects, induces a loss of size control, resulting in broader distributions of the added size at division, while keeping the median relatively constant. In terms of the division rate, we have shown, using a parametric Generalised Gamma model, that the adder division rate function $B$ is reduced by the SOS response in a nutrient-dependent way. In particular, the previously observed \textit{catastrophe} or decreasing regions in the division rate can be explained quantitatively by the contribution of high SOS individuals to division arrest. We observe however that the experimental heterogeneity of the joint SOS and size distributions is still more important than our model predicts. In this sense, one interesting axis of future work might be the extension of our model to include \textit{mixed effects} \cite{Lavielle2014}, that is, to allow individual heterogeneity in the parameters of the probability distributions of the model. For instance, this could enable having a Generalised Gamma model at the population level (the \textit{fixed effects}) consistent with the idea that the adder is robust \textit{in average}, but with some parameters that could show variability among individual cells (the \textit{random effects}). Further, this could also enable to statistically test the heterogenxeity of the population. Mixed Effects Models are extensively used to model individual-based responses in pharmacokinetics, for example, where the evolution of the drug in time is driven by a deterministic ODE. The extensions required to adapt the method to a stochastic diffusion process, such as our Ornstein-Uhlenbeck model, which is moreover coupled to the stochastic process of cell division, would constitute an interesting challenge see for example \cite{Donnet2008}.

\begin{acknowledgments}
This work has been supported by the Chair Modélisation Mathématique et Biodiversité of Veolia Environnement - École polytechnique - Museum National d'Histoire Naturelle - Fondation X and a Wellcome Investigator Award in Science to M.E.K. (Grant No. 205008/Z/16/Z). Funded by the European Union (ERC, SINGER, 101054787). Views and opinions expressed are those of the author(s) only and do not necessarily reflect those of the European Union or the European Research Council. Neither the European Union nor the granting authority can be held responsible for them. 
\end{acknowledgments}

\bibliography{biblio}

\appendix
\begin{widetext}
\section{Generalised Gamma model for the division rate}
\label{app:gengamma}
The division is supposed to be triggered by the added size since birth $a(t)$ (adder model) and the level of SOS intensity $X_t$. This is described by a division rate $\beta(a,y,X) = \lambda y B(a,X)$, where $\lambda$ is the elongation rate, supposed to be the same constant for all cells. Given $x$, $B(\cdot,x)$ is supposed to be the rate function of a Generalised Gamma distribution of parameters $(m_0, s(x), q(x))$. This means that under constant SOS expression $(X_t)_t \equiv x$,
    \[
     \left\{
                \begin{array}{ll}
                  \displaystyle \log A_{div} = m_0 + (s(x)/q(x)) \log \pars{q^2(x) R}  \\
                  R \sim \textrm{Gamma}(q^{-2}(x), 1) \\
                  s(x) = s_0 + s_1 x \; , \; q(x) = q_0 + q_1 x 
                \end{array}
    \right.
    \label{eq:GenGammaModelSummary}
    \]
    The intercepts $s_0$ and $q_0$ are to be obtained from the control dataset. The unknown parameters $s_1 > 0$ and $q_1 \in \rr$ measure the additional linear effect of the SOS level $x$ on the dispersion of the distribution around $m_0$. 

    \begin{figure}
        \centering
        \includegraphics[width=0.45\textwidth]{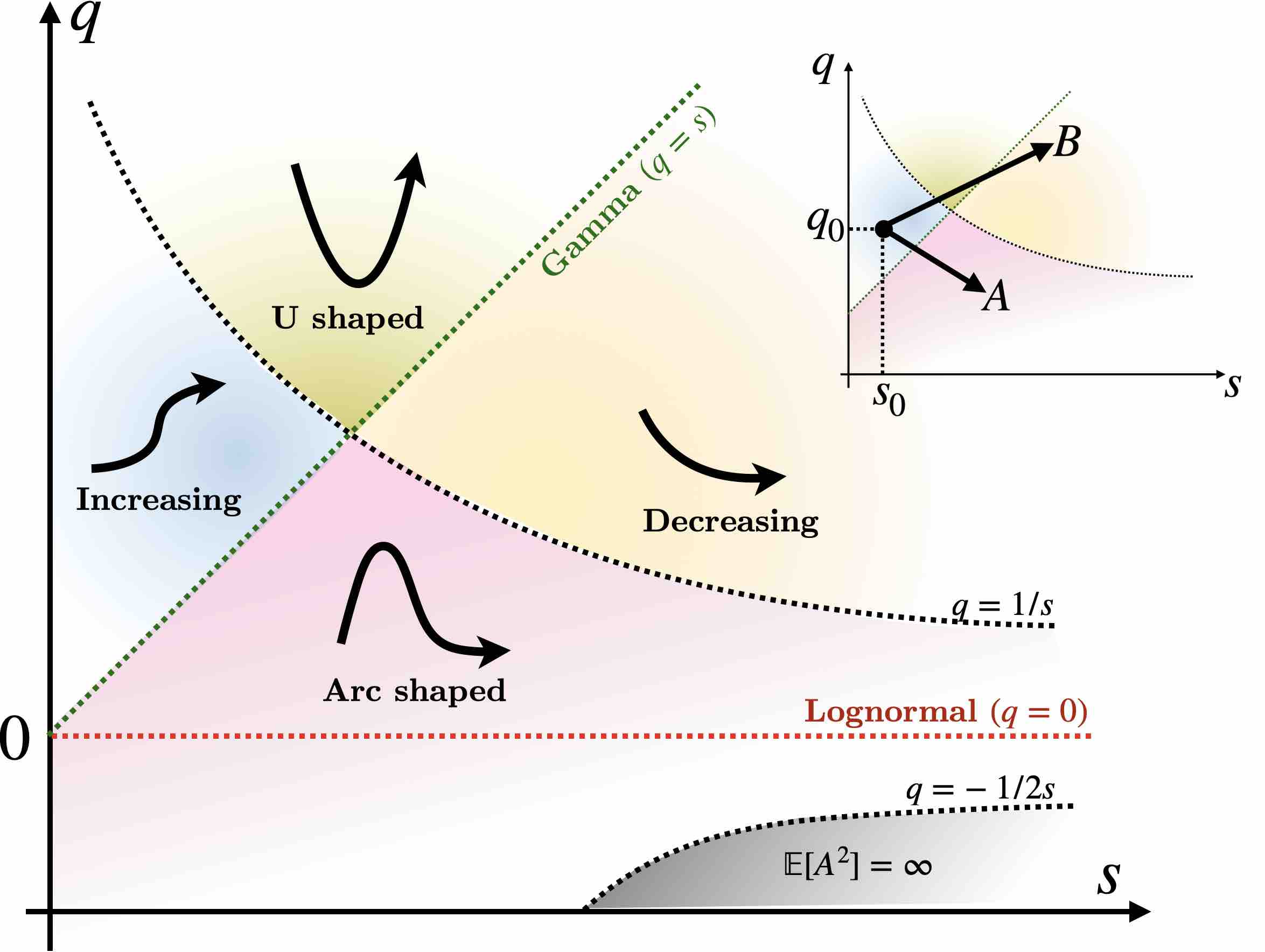}
        \caption{Adapted from \cite{Cox2007}. The 4 possible shapes of the division rate $B$ that can be obtained with a GenGamma($m,s,q$) model at fixed $m>0$. Inset: starting from $(s_0, q_0)$ giving a monotonically increasing division rate, and depending on the values of $q_1$ and $s_1$, an increased level of SOS response might lead to an arc shaped lognormal-like division rate ($A$), or to a monotonically decreasing division rate ($B$).}
        \label{fig:gengammas_cartoon}
    \end{figure}
    
 Cox et al. \cite{Cox2007} show that different values of $s$ and $q$ can generate very flexible rate functions (Fig. \ref{fig:gengammas_cartoon}). Indeed, the Gamma distribution of mean $e^{m}$ and coefficient of variation $s$ is obtained doing $q = s$. The Lognormal distribution, of log mean $m$ and log standard deviation $s$ is obtained by doing $q = 0$. This flexibility will be important to account for the SOS-induced filamentation, as our results will show further below.

    First, we suppose that the median added size (in log scale) does not depend on the intensity of the response. This means that we fix a medium-dependent constant $m(x) = m_0$ for all $x \in \rr$. Second, we make the following assumptions concerning the effect of SOS induction on the shape of the added size distribution. We make the strong assumption that both $q(x)$ and $s(x)$ are affine functions of the SOS level $x$. This is, we introduce two parameters $q_1 \in \rr$ and $s_1 > 0$ such that
    \begin{align*}
    q(x) &= q_0 + q_1 x, \\
    s(x) &= s_0 + s_1 x,
    \end{align*}
    where the intercepts $q_0$ and $s_0$ are obtained in absence of SOS response and will be supposed known. Then, starting from a certain $(s_0, q_0)$, depending on the values of $q_1$ and $s_1$, the value of $x$ can change the shape of the division rate as shown by the example at the left panel of Fig. \ref{fig:gengammas_cartoon}.

\section{Parameter estimation of the coupled SOS-adder dynamics using cell lineages}
\label{app:MLE}

We model our observations as a discrete sample of $I$ independent realisations during time $t \in [0,T]$, with $T = N \Delta t$. This is, we consider a sample 
\[
\pars{ O^{i}_{n \Delta t} }_{n=0,...,N}^{i=1,...,I}= (A^{i}_{n \Delta t}, Y^{i}_{n \Delta t}, X^{i}_{n \Delta t})_{n=0,...,N}^{i=1,...,I}
\]

To model this discrete-time process we define first $p(a,y,x)$ as the probability to divide in the following $\Delta t$ interval starting with state $(a,y,x)$. Then, for all $n$ we let $U_n^i$ be a Bernoulli random variable of parameter $p(A_n^i, Y_n^i, X_n^i)$, this is, which is equal to 1 if the lineage $i$ divides in the interval $[n \Delta t, (n+1) \Delta t[$, and $0$ otherwise. Since the intervals of time when a a division occurs are observed, $U_n$ is also an observed variable, available from the mother machine data. 

Notice that the probability $p(a,y,x)$ is given by
\begin{align*}
p(a,y,x) &= \mathbb{E}_{x} \left[ \int_0^{\Delta t} \beta(a + y(e^{\lambda t} - 1),y e^{\lambda t},X_t)  
\exp \pars{ - \int_0^t \beta(a + y(e^{\lambda s} - 1),y e^{\lambda s},X_s)  \d s } \d t  \right] 
\end{align*}

However, since $\Delta t$ is small enough, we do the first-order approximation $
p(a,y,x) \approx \beta(a,y,x) \Delta t $. Thus, given $(A_0^i, Y_0^i, X_0^i )$ we can generate $(A_n^i, Y_n^i, X_n^i )$ by the following hierarchical model (since they are i.i.d. samples, we forget the $i\in \setof{1,...,I}$ corresponding to each independent lineage):
\begin{align}
\textrm{Draw independently } & 
\begin{cases}
    U_{n} \sim& \textrm{Bernoulli}(p(A_n,Y_n,X_n)) \\
\rho_n \sim&  k(Y_n e^{\lambda \Delta t},\cdot)  \\
W_n \sim& \mathcal{N}(0,1) 
\end{cases} \\
c_n &= \mathds{1}_{2 < n \Delta t \leq 14} \\
X_{n+1} &= X_{n} e^{-\theta_{c_n} \Delta t} + \mu_{c_n} \pars{1 - e^{-\theta_{c_n} \Delta t}} + \zeta_{c_n} \sqrt{(1-e^{-2\theta_{c_n} \Delta t})/(2 \theta_{c_n})} W_n
\label{eq:OUsolution} \\
Y_{n+1} &= U_n \rho_n Y_n e^{\lambda \Delta t} + (1-U_n) Y_n e^{\lambda \Delta t} \label{eq:sizesJump} \\
A_{n+1}  &= (1-U_n) (A_n + Y_{n+1} - Y_n) \label{eq:agesJump} 
\end{align}

Eq. \eqref{eq:agesJump} resets the added size at 0 at each division (i.e., when $U_n = 1$), and otherwise adds the increment of size $Y_{n+1}-Y_{n}$, with $Y_{n+1}$ given by Eq. \eqref{eq:sizesJump}. When a division occurs the size is multiplied by the daughter-to-mother size ratio $\rho_n$ distributed according the size-dependent probability kernel $k(y, \cdot)$ defined in \eqref{eq:kSOS}. Eq. \eqref{eq:OUsolution} corresponds to the explicit solution of the Ornstein-Uhlenbeck Equation \eqref{eq:OU}.

\subsection{Likelihood of the observations}
Let $\eta = (\eta_X, \eta_\beta, \eta_k)$ the vector of parameters considered. From the previous set of equations, the log-likelihood of the observations under the considered parametric model is given by

\begin{align}
\log \mathcal{L}((A_n^i,Y_n^i,X_n^i,U_n^i) | \eta)  := & \sum_{i \geq 1, n \geq 1} 
 \log \mathbb{P}((A_n^i,Y_n^i,X_n^i,U_n^i) , (A_{n+1}^i,Y_{n+1}^i,X_{n+1}^i,U_{n+1}^i) | \eta) \nonumber \\
= & \; \ell_1 ((Y_n^i,U_n^i) | \eta_k)  \nonumber + \ell_2 ((A_n^i,Y_n^i,X_n^i,U_n^i) | \eta_\beta )  
+ \ell_3 ((X_n^i) | \eta_X) \nonumber \\
& + \textrm{constant independent from $\eta$} \nonumber
\end{align}
where
\begin{align}
    \ell_1 ((Y_n^i,U_n^i) | \eta_k )
    = \sum_{i=1}^I  & \  \sum_{M=1}^{+\infty} \ \sum_{ \substack{ n \in \llbracket 0, N \rrbracket : \\ Y_{n-1}^i \leq 2My^*, \\  Y_{n-1}^i > 2(M-1)y^* }}  U_n^i \log \pars{ \sum_{m=1}^{M} \frac{1}{M} F^M_m \pars{\left. \frac{Y_n^i}{Y_{n-1}^i e^{\lambda \Delta t}} \right| v }  },
    \label{eq:ell1} \\
    \ell_2 ((A_n^i,Y_n^i,X_n^i,U_n^i) | \eta_\beta )
    = \sum_{i=1}^I  & \left( \sum_{n=1}^{N}  U_n^i \log \pars{ \beta(A_n^i,Y_n^i,X_n^i | \eta_\beta) \Delta t }  \right. \nonumber \\
    & + \left. \sum_{n=1}^{N}  (1-U_n^i) \log \pars{ 1- \beta(A_n^i,Y_n^i,X_n^i | \eta_\beta ) \Delta t } \right)
    \label{eq:ell2} \\
    \ell_3 ((X_n^i) | \eta_X) =  \sum_{i=1}^I & \sum_{n=0}^{N-1} \log  \mathfrak{g} \pars{ X_{n+1}^i  \bigg|  X_{n} e^{-\theta_{c_n} \Delta t} + \mu_{c_n} \pars{1 - e^{-\theta_{c_n} \Delta t}}  , \;  \frac{\zeta_{c_n}^2 (1-e^{-2\theta_{c_n} \Delta t})}{2 \theta_{c_n}}  }
    \label{eq:ell3}
\end{align}
and where $\mathfrak{g}(\cdot|\mu; \sigma^2)$ is the Gaussian distribution of mean $\mu$ and variance $\sigma^2$ and $k$ is given by Eq. \eqref{eq:kSOS}, parameterised by $y^*$ and $c$. The division probability $p$, which depends on $\beta$, is parameterised by $v_1$.  In particular the contributions of the parameters related to $k$, to $\beta$, and to the Ornstein-Uhlenbeck are all independent. 

\begin{figure*}[t]
    \centering
    \includegraphics[width=0.75\textwidth]{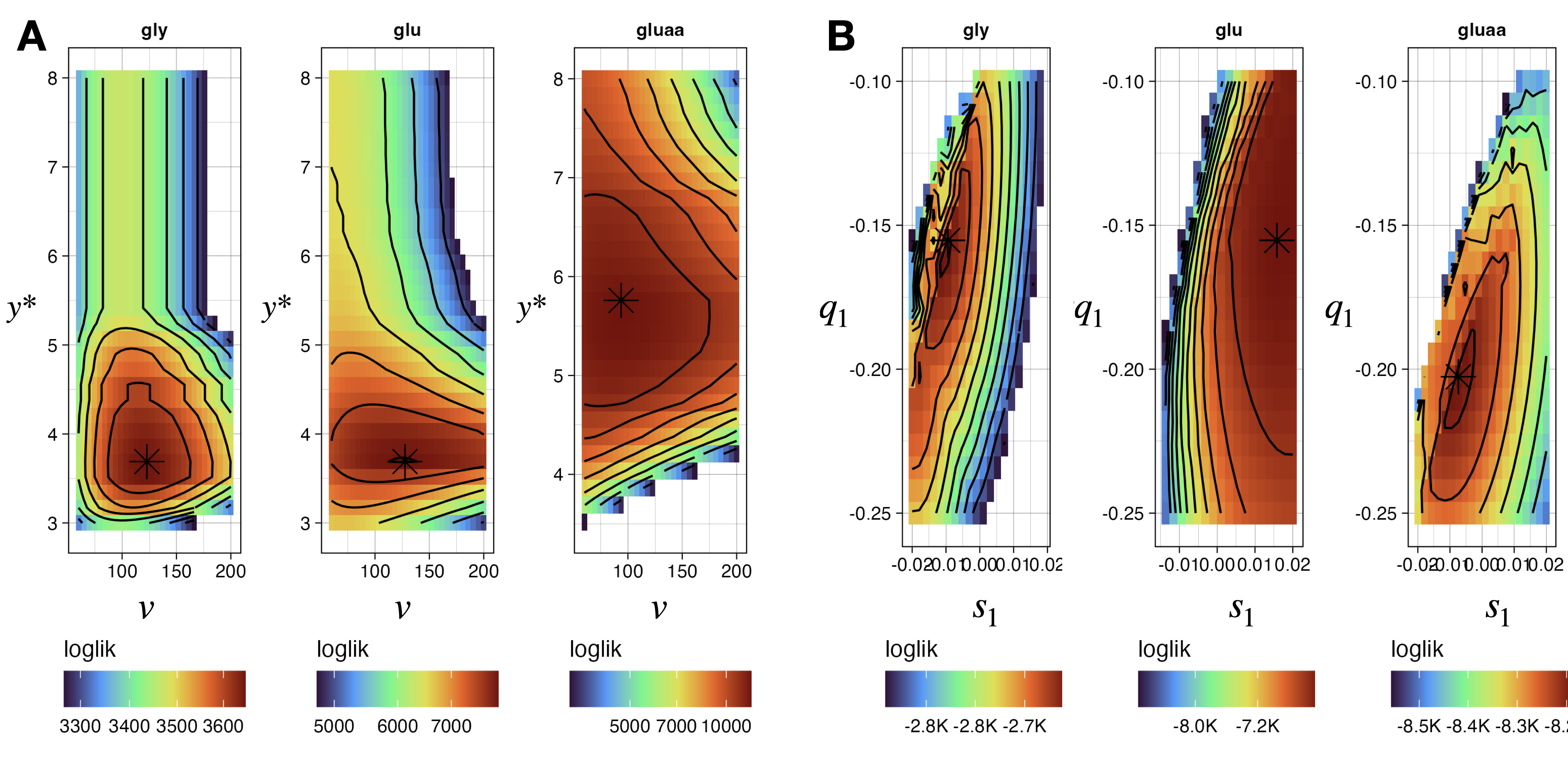}
    \caption{\textbf{Log-likelihoods of the parameters of the SOS perturbed adder model.} The MLE are marked $*$. \textbf{A.} Log-likelihood $\ell_1$ \eqref{eq:ell1} of the division size observations as function of the critical mother length $y^*$ and the concentration parameter $v$ of the Beta mixture \eqref{eq:kSOS}-\eqref{eq:kSOS_Beta}. \textbf{B.} Log-likelihood $\ell_2$ \eqref{eq:ell2} of the added sizes at division as function of $s_1$ and $q_1$.}
    \label{fig:likelihood_F_B}
\end{figure*}

We see that the likelihoods of the model can be computed explicitly, and we show below some first numerical results concerning their computation. One of the remarkable properties of the Ornstein-Uhlenbeck process is that the three parameters $(\theta, \mu, \zeta)$ possess explicit Maximum Likelihood Estimators (MLE) \cite{GenonCours, Tang2009}. Thus, using the data from the time interval $t \in ]2, 14]$, in which the cells are under the effect of the antibiotic, we infer the values of of $(\theta_1, \mu_1, \zeta_1)$. Using the remaining time (pre and post exposure), we infer the values of $(\theta_0, \mu_0, \zeta_0)$.  
Contrary to the Ornstein-Uhlenbeck process, the likelihoods $\ell_1$ and $\ell_2$ do not allow to obtain the MLE in close forms. However, $\ell_1$ and $\ell_2$ are both numerically tractable and our computations show that they are convex (see Fig. \ref{fig:likelihood_F_B}), so that the numerical maximisation can be done by classical approaches.  Fig. \ref{fig:likelihood_F_B}A gives the value of $\ell_1$ as functions of $y^*$ and $v$. We see that the log-likelihood has convex contour levels, and a unique global maximum. Similarly, we can compute the value of $\ell_2$ \eqref{eq:ell2}. Fig. \ref{fig:likelihood_F_B}B shows the log-likelihood as function of the SOS-induced linear factors $s_1$ and $q_1$ multiplying the dispersion parameters of the Generalised Gamma model \eqref{eq:GenGammaModelSummary}. The intercepts $s_0$ and $q_0$ were inferred as the MLE of a Generalised Gamma fitted directly to the added size distributions of the 10\% of cells with lowest SOS signal at division. The numerical computations show that $\ell_2$ also has convex contours and a unique global maximum. All the inferred parameters are tabulated in Table \ref{tab:Division_MLEvalues}.

\section{Medium-dependent division statistics on control dataset}
\label{app:fitF_ctrl}

We fit a Beta distribution $\hat F$ as a suitable estimator of the distribution of the ratio $\rho$ observed in datasets obtained absence of ciprofloxacin for the three media \cite{Broughton2024}. The estimated parameters along with their 95\% confidence intervals are given in Table \ref{tab:paramsF} for the three different media. The bigger the value of $\alpha$ and $\beta$ the more concentrated the distribution, which can be observed in Fig. \ref{fig:fitF}. We see that the distribution is wider in richer media. This could mean that the position of the division septum is less exact in fast growing bacteria. 
	\begin{table}[h]
		\centering
\begin{tabular}{|l|l|l|}
\hline
Medium & $\alpha$ & $\beta$ \\
\hline
Glycerol & $20.5415 \pm 0.69271$ & $20.5410 \pm 0.69269$ \\
\hline
Glucose & $8.9463 \pm 0.59662$ & $8.9466 \pm 0.59664$ \\
\hline
Glucose+aa & $5.9614 \pm 0.14157$ & $5.9617 \pm 0.1415835$ \\
\hline
\end{tabular}
\caption{Beta distribution parameters and 95\% confidence intervals for the fitted mother-to-daughter ratio in each medium.}
\label{tab:paramsF}
	\end{table}

		\begin{figure}[h]
		\centering
		\includegraphics[width=0.45\textwidth]{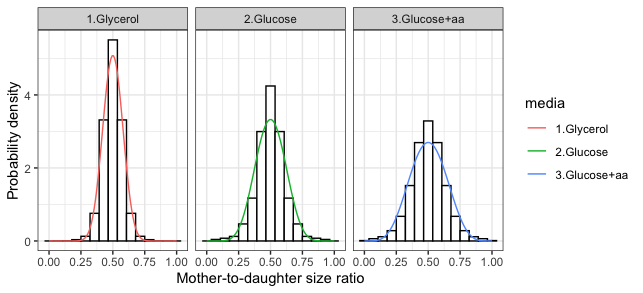}
		\caption{Estimated densities $\hat F$ for the three different media according to the parameters of Table \ref{tab:paramsF}}
		\label{fig:fitF}
	\end{figure}

 \end{widetext}
 
\end{document}